\documentclass[12pt,a4paper,dvips]{article}

\usepackage{epsfig} 

\renewcommand\thesection{\arabic{section}.}
\renewcommand\thesubsection{\thesection\arabic{subsection}.}
\renewcommand\thesubsubsection{\thesubsection\arabic{subsubsection}.}
\renewcommand\section[1]{\vspace{\topsep}\vspace{\partopsep}
\refstepcounter{section}
{\par  \noindent\normalsize\bfseries \thesection
\hspace{1em}#1\vspace{\topsep}\par\noindent}}

\newenvironment{refs}
{\vspace{\topsep}\vspace{\partopsep}
{\par \noindent\normalsize\bfseries  References
\vspace{-\topsep}\par\noindent}
\setlength{\parindent}{-5mm}
\begin{list}{}{\topsep 0pt \partopsep 0pt \itemsep 0pt \leftmargin 5mm
\parsep 0pt \itemindent -5mm}}
{\end{list}}

\renewcommand\subsection[1]{
\refstepcounter{subsection}
{\par \protect\vspace{\topsep}\vspace{\partopsep}
 \noindent\normalsize\bfseries \slshape \thesubsection
\hspace{1em}#1\par \noindent}}

\renewcommand\subsubsection[1]{
\refstepcounter{subsubsection}
{\par \protect \vspace{\topsep}\vspace{\partopsep}
\noindent\normalsize \slshape \thesubsubsection
\hspace{1em}#1\par \noindent}}

\newfont{\sansb}{cmssbx10}
\newfont{\sans}{cmss10}
\begin{document}
\begin{center}
{\large \bf Measuring $\gamma$-Ray Energy Spectra 
with the HEGRA IACT System 
\footnote{Talk presented at the Workshop ``Towards a Major Atmospheric
Cherenkov Detector V'', Kruger Park, South Africa, 1997}
\vspace{18pt}\\}
{W. Hofmann \\ HEGRA Collaboration \vspace{12pt}\\}
{\sl 
Max-Planck-Institut f\"ur Kernphysik, Heidelberg \vspace{-12pt}\\
}
\end{center}

\begin{abstract}
The stereoscopic reconstruction of air showers viewed by
multiple imaging atmospheric Cherenkov telescopes (IACTs) allows a
more precise reconstruction of shower energies and hence an
improved determination of energy spectra. Reconstruction
techniques and in particular new systematic checks are 
discussed on the basis of the large sample of $\gamma$-rays
from Mkn 501 detected with the HEGRA IACT system.
\end{abstract}
\setlength{\parindent}{1cm}
\section{Introduction}
\begin{figure}[hb]
\begin{center}
\epsfig{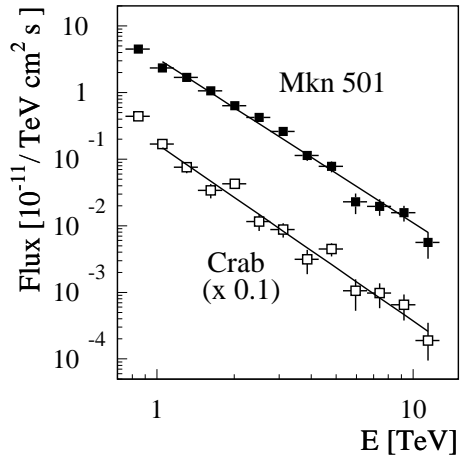}
\end{center}
\caption
{Typical energy spectra measured using the HEGRA
telescope system. Errors are statistical only.}
\label{fig_spectra}
\end{figure}

One of the premises of stereoscopic imaging of air showers using 
multiple atmospheric Cherenkov telescopes is an improved reconstruction
of shower energies and hence of source spectra. Stereoscopic imaging
allows a rather precise reconstruction of the core location; using 
the known relation between light yield, core distance, and shower energy,
the energy is estimated by suitably averaging the measurements by the
individual telescopes.
The HEGRA system of imaging atmospheric Cherenkov telescopes 
(Daum et al., 1997) is
operational 
since late 1996, and first results on energy spectra have been shown
at various occasions; for example, Fig.~\ref{fig_spectra} shows the
current version of the Mkn 501 spectrum (Aharonian et al., 1997). 
Purpose of the present paper is
to provide background information on the techniques used to reconstruct
spectra, and to discuss the status of our understanding of the 
system and the systematic checks of the energy reconstruction, which are
possible for the first time due to the redundant information provided by
the telescope system. I will very briefly discuss the selection of 
$\gamma$-ray events and then concentrate on two areas, the energy estimate
and the determination of the effective detection area. I should 
emphasize that the primary goal in these analyses was 
to minimize systematic uncertainties, sometimes at the expense of
some loss in statistical precision. 

\setlength{\parindent}{1cm}
\section{$\gamma$-ray event selection}
\begin{figure}
\begin{center}
\epsfig{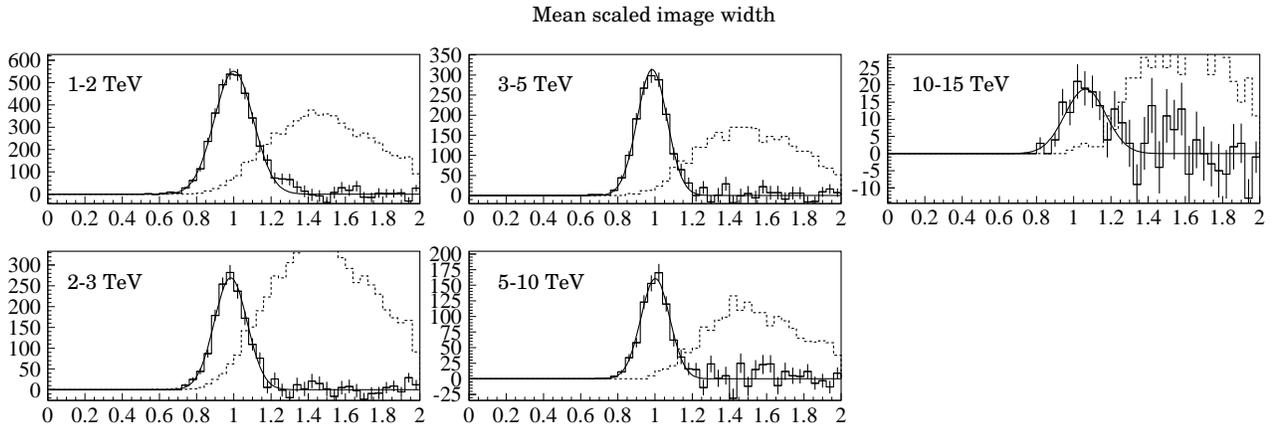}
\end{center}
\caption
{Mean scaled width of events in different ranges of reconstructed energy.
The full line shows the distribution of excess events from the direction
of Mkn 501 (and a Gaussian fit), 
the dashed line indicates the off-source background.}
\label{fig_msw}
\end{figure}

\begin{figure}
\begin{center}
\epsfig{file=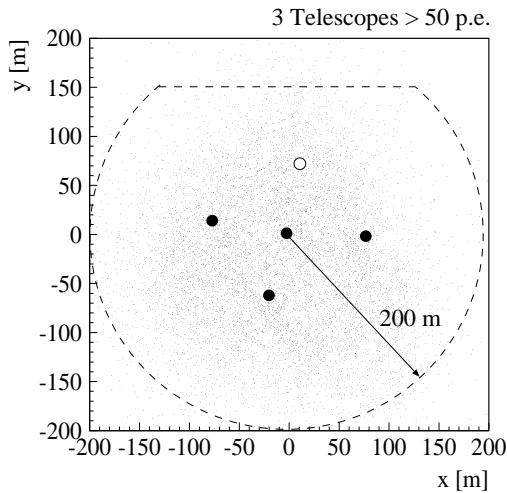,height=6.5cm,angle=0}
\end{center}
\caption
{Distribution of reconstructed shower cores, together with the locations 
of the telescopes. Telescope CT2 (open circle) is not yet included in
the IACT system. The dashed lines indicates the acceptance region.}
\label{fig_core}
\end{figure}

For a $\gamma$-ray source such as Mkn 501, events are selected on the
basis of their direction, of the event shape, and of the core location.
The directional cuts and the estimate of their efficiency is 
relatively straight forward. 
Next, $\gamma$-rays are enriched using the shape information. The so called
{\em mean scaled width} is calculated by determining for each telescope,
as a function of image {\em size}, distance to the core and zenith angle the
expected {\it width} of the image, normalizing the measured {\em width} to the
expected {\it width}, and averaging over all telescopes. Fig.~\ref{fig_msw}
shows the mean scaled width for different ranges of the (reconstructed)
shower energy. Excess events from the direction of Mkn 501 show up as a 
peak around a mean scaled width of 1; the cosmic-ray background gives rise
to a broad distribution at higher width values. In the determination of 
spectra, typical cuts require a mean scaled width below 1.3, resulting in an
efficiency well above 90\%. 
As an additional cut, the location of the shower core is restricted. 
Fig.~\ref{fig_core} illustrates the distribution of shower cores as well
as the locations of the four telescopes presently included in the IACT
system. Event are required to lie within 200~m from the central telescope, 
CT3. Usually, we also remove events with a $y$ coordinate greater than
150~m, reflecting the fact that the telescope CT2 is not yet included
in the CT system. This selection
guarantees that a least one telescope is within about 150~m from the shower
core.

\section{Determination of core location}
\begin{figure}
\begin{center}
\epsfig{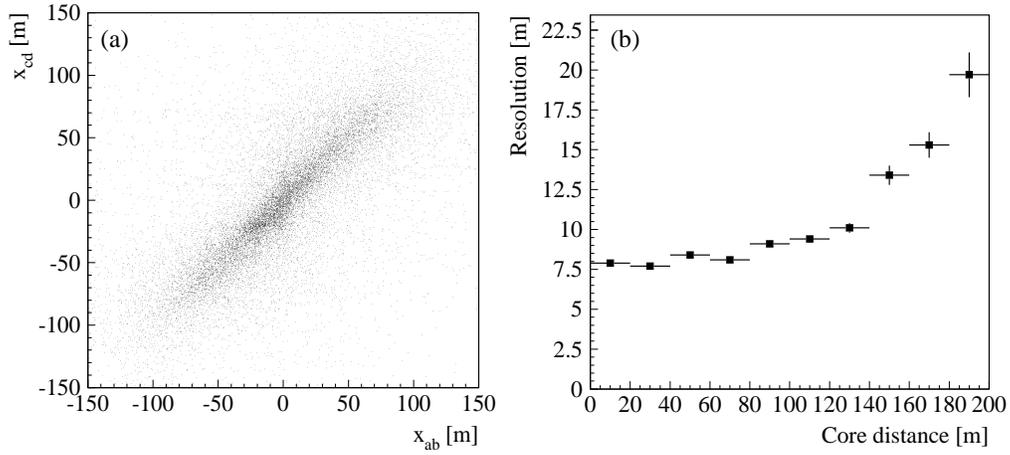}
\end{center}
\caption
{(a) Scatter plot of $x$-coordinate of the core 
measured with one pair of telescopes,
vs that measured with the other pair, for $\gamma$-rays from Mkn 501.
(b) Error in the determination of the $x$-coordinate of the shower core
for 4-telescope events as a function of core distance from the
central telescope, derived assuming that the two 2-telescope
measurements are independent. The resolution given refers to a Gaussian fit
to the distributions; there are slight non-Gaussian tails.}
\label{fig_core2}
\end{figure}

A crucial input both for the event selection and the energy reconstruction
is the reconstruction of the location of the shower core. The reconstruction
is based on simple geometry (Daum et al., 1997). With a 4-telescope system, it 
is not only possible to reconstruct shower cores, but one can also test the
precision of the core reconstruction by dividing the full system into two
groups of two telescopes each, reconstructing the core separately for each
group and comparing the results. In Fig.~\ref{fig_core2}(a), the $x$-coordinate
obtained from one group is plotted vs the $x$-coordinate
 from the other group. One sees
a very clear correlation. Assuming that both measurements are uncorrelated,
the error in the determination of the core location for such 4-telescope events
can be determined (Fig.~\ref{fig_core2}(b)). The error in $x$ (and also in $y$)
is about 8~m for events with the core in the central region of the array, and 
increases to 20~m for distant events, where fewer telescopes trigger and 
where stereo angles are small. 

\section{Energy determination}
\begin{figure}
\begin{center}
\epsfig{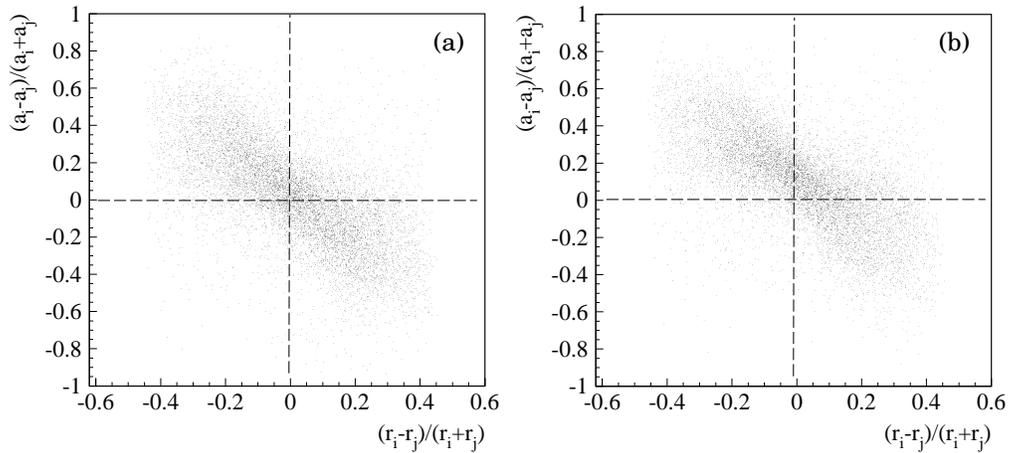}
\end{center}
\caption
{Asymmetry in measured amplitudes $a_i, a_j$ for a pair of telescopes
$i, j$, 
as a function of the asymmetry in the distances $r_i, r_j$ from the
telescopes to the shower core. (a) For CT3 and CT4, (b) for CT3 and CT6.}
\label{fig_relcal}
\end{figure}

As a first step in the reconstruction of shower energies, the response 
of the telescopes needs to be calibrated. We use a two-step approach, 
first equalizing the relative response of the telescopes, and then 
adjusting the global scale. The determine the relative sensitivity 
of two telescopes $i$ and $j$, shower cores are reconstructed using these two 
telescopes only. Then, the asymmetry in image sizes, 
$(a_i-a_j)/(a_i+a_j)$ is plotted vs the asymmetry in the distances
from the core to the telescopes, $(r_i-r_j)/(r_i+r_j)$ 
(Fig.~\ref{fig_relcal}). If two telescopes have a similar sensitivity,
the average asymmetry in image sizes is zero for $r_i \approx r_j$
(Fig.~\ref{fig_relcal}(a)). In the telescopes are unequal, the 
distribution appears shifted (Fig.~\ref{fig_relcal}(b)). We find the 
following ratios of sensitivities: CT4/CT3 = 0.937 (0.937),
CT5/CT3 =  0.857 (0.855), and CT6/CT3 = 0.791 (0.778). The first
number is always determined using a cosmic-ray event sample, the
second (in parenthesis) using the Mkn 501 $\gamma$-ray sample.
The two sets of numbers are consistent within their statistical
errors, which range from 0.005 to 0.007. We believe, therefore, that
the relative calibration is good to about 1\% or better. The 20\%
differences between telescopes could be caused by differences in
the quantum efficiencies of the different batches of PMTs, by
variations in mirror reflectivity etc. (the telescopes came
into operation sequentially, over a span of 1.5 years).

The absolute calibration of the energy scale is much more challenging.
Presently, we study three different methods: i) the determination of
the photoelectron to ADC channels conversion factor using the width of
the distribution of laser test pulses to estimate the mean number
of photoelectrons per pulse, ii) the measurement of the response
of the whole telescope and of the analysis chain simultaneously using
a distant, pulsed, calibrated light source (Frass et al., 1997), and 
iii) the calibration by comparing the measured and the expected 
cosmic-ray rate (A. Konopelko et al., 1996). For example, at a fixed reference
wavelength of 430~nm the first method results in 
a sensitivity of $(1.13 \pm 0.25) \cdot
10^4$ ADC channels/(photon/cm$^2$), the second in  $(1.20 \pm 0.12) \cdot
10^4$ ADC channels/(photon/cm$^2$). For the cosmic-ray comparison,
the MC is still being refined, but first analyses indicate values
consistent with the other two, with errors of about 15\%.
Ultimately, we believe that the energy scale can be determined
within about 10\%. Note that all energy
spectra shown so far use the energy scale derived from a previous 
generation of MC simulations, where the detector simulation had known
imperfections; scales are likely to change by 10\% to 20\%.

\begin{figure}
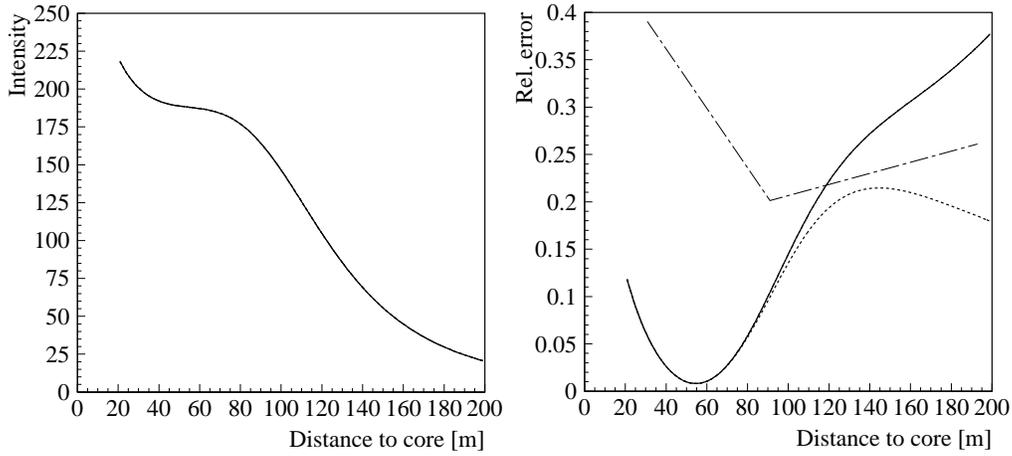

\begin{center}
\epsfig{file=amp_vs_r.eps,height=6cm,angle=0}
\epsfig{file=damp_vs_r.eps,height=6cm,angle=0}
\end{center}
\caption
{(a) (Left) Typical relation between mean light yield and distance to the 
shower core, for TeV $\gamma$-ray showers. (b) (Right) Relative error of a 
single-telescope energy estimate as a function of core distance,
as assumed in the energy reconstruction. 
Dashed-dotted: shower fluctuations and photoelectron statistics; dotted line:
contribution from the error in the distance measurement for a constant
error in core position; full line: for a core error as shown in Fig. 4(b).}
\label{fig_sizecore}
\end{figure}

Once the sensitivity is known (or assumed), the shower energy is
determined by converting the image size seen in each telescope 
into an energy value, using the measured distance to the core and
the relation between energy, core distance, zenith angle and image size obtained
from simulations. Two different approaches are used, one based on 
an analytical approximation, the other based on Monte-Carlo generated
look-up tables. Fig.~\ref{fig_sizecore}(a) illustrates a typical dependence
of average image size on core distance. To properly average over telescopes,
the errors of the energy estimates need to be known. One contribution are
shower fluctuations and photoelectron statistics (Fig.~\ref{fig_sizecore}(b));
these errors are large close to the core, and roughly constant at about 20\%
for larger core distances. A second contribution arises from the error in
the determination of the shower core; this contribution is proportional 
to the slope of the size-distance relation, and to the core error, and
rises steeply for core distances of 100~m and above.

\begin{figure}
\begin{center}
\epsfig{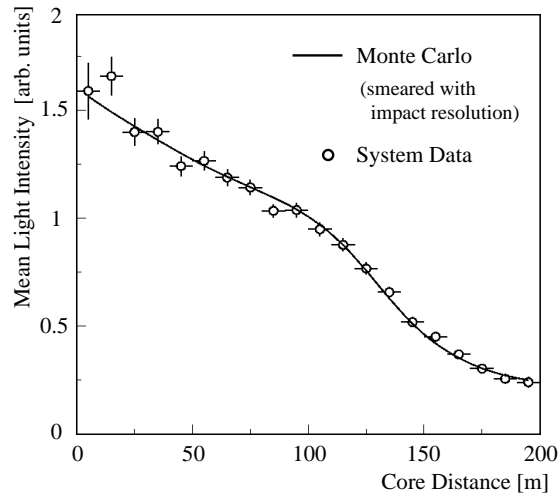}
\end{center}
\caption
{Measured mean image size as a function of core distance, for showers of
fixed energy, using the Mkn 501 data sample (see text for details). 
The line shows the result of
Monte Carlo simulations, smeared with the experimental resolution
in the core distance.}
\label{fig_ampvsdist}
\end{figure}

A new and unique feature of the telescope systems is that these crucial
inputs can be verified experimentally. To measure the light yield as a 
function of core distance, one selects events with cores at a 
fixed distance from telescope $i$, and with a fixed image size $a_i$. These
conditions select showers with a (roughly) fixed energy. One then plots
for one of the other telescopes $j$ the image size as a function of the
(varying) core distance $r_j$. Of course, care has to be taken not to bias
the results due to trigger conditions and selection criteria.
Fig.~\ref{fig_ampvsdist} shows that the measured distributions agree 
reasonably with the simulations.

\begin{figure}
\begin{center}
\epsfig{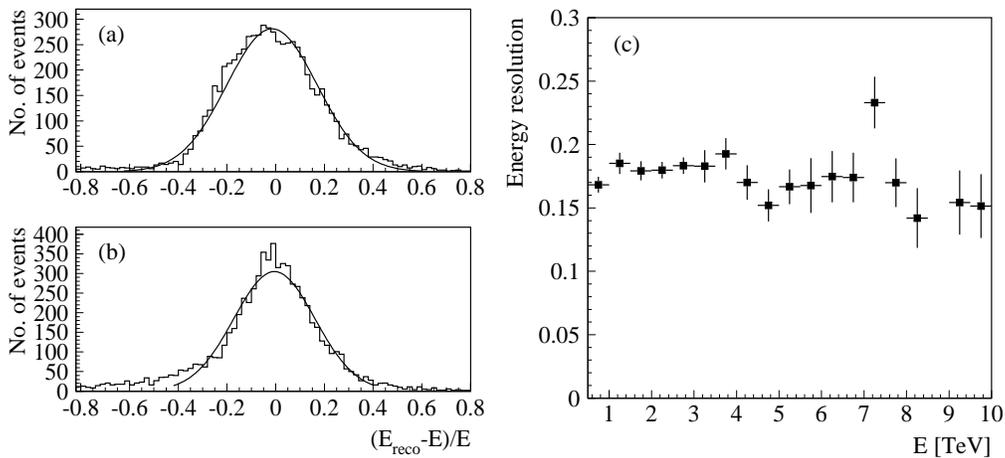}
\end{center}
\caption
{Monte-Carlo studies of the energy resolution. (a,b) Distribution in the
relative difference between reconstructed and true energies, for two
algorithms.(c) Energy resolution as a function of $\gamma$-ray energy.}
\label{fig_eres}
\end{figure}

\begin{figure}
\begin{center}
\epsfig{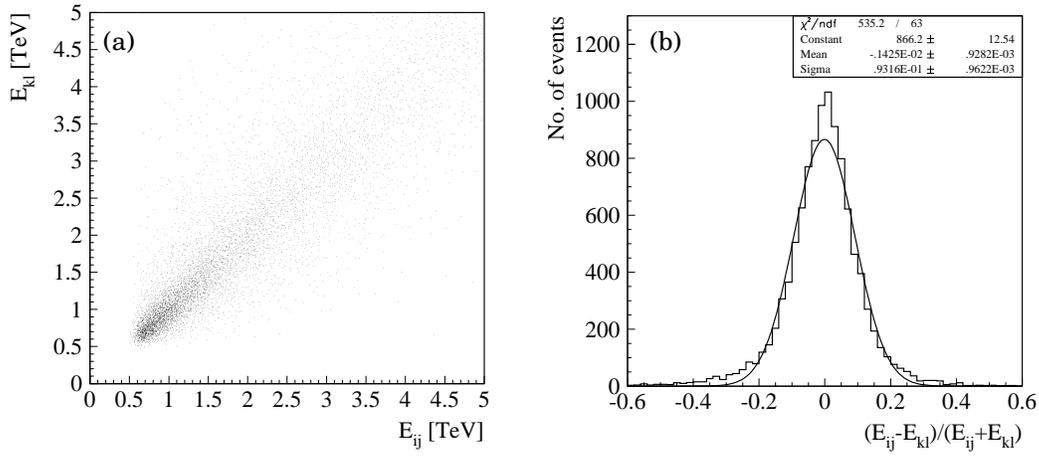}
\end{center}
\caption
{(a) Scatter plot of shower energy 
measured with one pair of telescopes,
vs. that measured with the other pair, for $\gamma$-rays from Mkn 501.
(b) Relative difference in the energy values measured by 
two pairs of telescopes.}
\label{fig_expe}
\end{figure}

The shower energy is derived by averaging over the measurements from the
individual telescopes. Monte Carlo studies show that the resulting energy
resolution is not sensitive the details of the averaging process, such as
the exact choice of the weights, or if $\log{E}$ instead of $E$ is averaged.
Figs.~\ref{fig_eres}(a,b) shows energy resolutions slightly below 20\% obtained
with two different algorithms. Both show a non-Gaussian tail towards low
reconstructed energies, caused by events where the core location is
poorly determined. The tail, however, does
not harm the reconstruction of steeply falling spectra.
The energy resolution is almost independent of energy over a wide range
(Fig.~\ref{fig_eres}(c)).
Again for the first time the IACT system allows to check the energy 
reconstruction procedure by determining energies of 4-telescope events
independently from two pairs of two telescopes each. Fig.~\ref{fig_expe}(a)
shows that the two measurements correlate quite well. Assuming that the
two measurements are independent, one can derive the energy resolution
of the system from the width of the distribution in the difference of the
two measurements (Fig.~\ref{fig_expe}(b)). The resulting resolution figure
of 9.3\% is significantly better than the resolution expected on the basis
of simulations, which seems to indicate that the two measurements are
correlated. Such a correlation might e.g. be caused by the variation in the
height of the shower maximum; if this is indeed the case, it should be 
possible to improve the resolution over the values given in 
Fig.~\ref{fig_eres},
since the height of the shower maximum can be reconstructed with a resolution
of about 1 radiation length (M. Ulrich et al., 1997).

\vspace{1cm}
\section{Effective detection area}
\begin{figure}
\begin{center}
\epsfig{file=area_k.eps,height=6.5cm,angle=0}
\epsfig{file=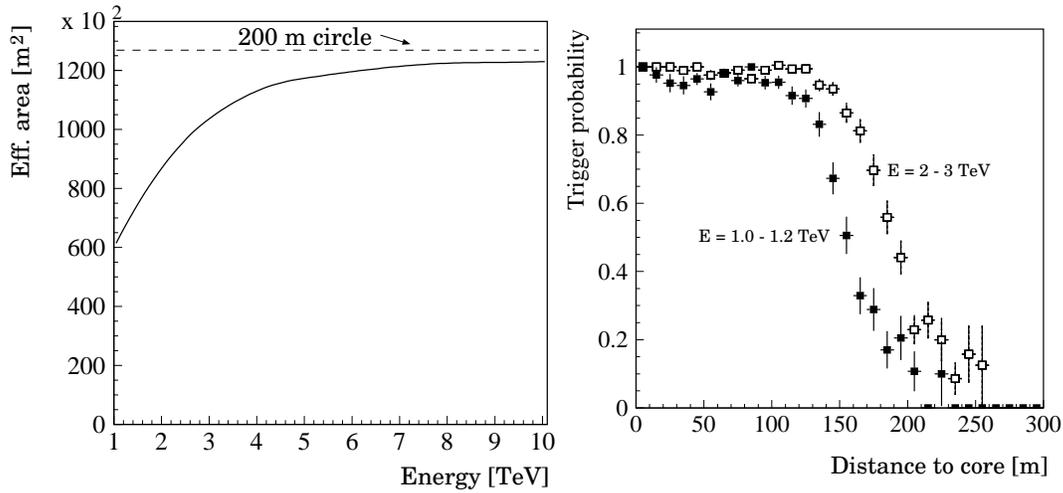,height=6.2cm,angle=0}
\end{center}
\caption
{(a) (Left) Effective detection area determined from simulations, as a function of
energy. Only events within 200~m from the central telescope are accepted
(b) (Right) Measured single-telescope trigger efficiency for 
showers in the 1.0 - 1.2 TeV and 2 - 3 TeV
energy range, as a function of the distance between the telescope and 
the shower core.}
\label{fig_area}
\end{figure}

Different techniques are used to determine effective areas, with consistent
results. In one approach, the effective area is determined from simulations
in the usual way, imposing an upper limit on the distance of the core
from the central telescope of 200~m. The resulting area (Fig.~\ref{fig_area}(a))
saturates at $\pi$(200~m)$^2$ for energies above a few TeV. Alternatively,
the simulations are used to define an energy-dependent maximum radius, such
that for events within this radius the average trigger efficiency if large 
($> 80\%$). The resulting radius is about 100~m at 1~TeV and rises to
200~m at about 4~TeV; above this energy, a constant 200~m cut is applied.
To determine spectra, in this latter case a (trigger-)efficiency correction
of $(10 \pm 10) \%$ is applied.

All these estimates of effective areas rely on the ability of the simulation
to reproduce single-telescope trigger efficiencies. These, however, can
in turn be determined from system data. One reconstructs an $n$-telescope
event using only a subset of $n-1$ telescopes, and then checks if the 
remaining telescope has triggered. Fig.~\ref{fig_area}(b) shows the 
measured trigger
efficiency obtained for showers in two energy ranges as a 
function of core distance. At $E \approx 1$~TeV,
the trigger 
efficiency saturates at $> 90\%$ out to distances of 120~m. 
With this measured dependence,
one can, e.g., verify that
TeV showers with cores within 100~m from the central telescope will 
indeed trigger two telescopes with more than 80\% average probability.

\begin{figure}
\begin{center}
\epsfig{file=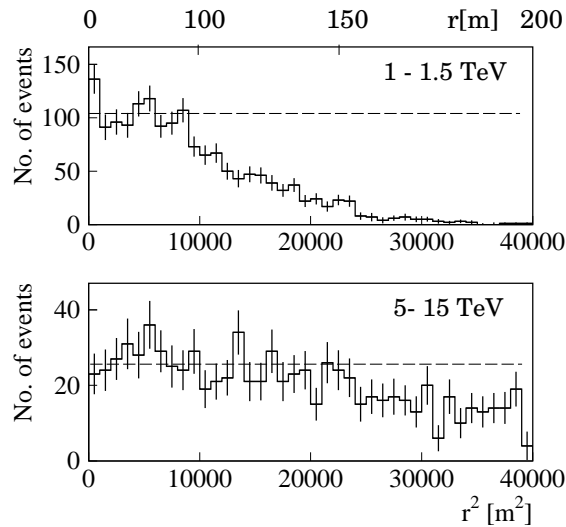,height=7cm,angle=0}
\end{center}
\caption
{Distribution $dn/dr^2$ of shower cores relative to the central
telescope, for two energy ranges.}
\label{fig_r2}
\end{figure}

Another way to verify the estimate the maximum radius for efficient triggering
of the IACT system is to simply plot the distribution $dn/dr^2$ of events,
where $r$ is the distance to the central telescope.
Ideally, the area density $dn/dr^2$ should be flat up to the 
(energy-dependent) maximum radius. Fig.~\ref{fig_r2} shows this distribution
for energies of 1 TeV to 1.5 TeV and for high energies above 5~TeV. The events above 5~TeV
show a slight drop-off beyond radii of 150~m; however, the trigger
efficiency averaged 
out to 200~m radius is still 84\% (assuming that is it indeed 100\%
for small radii). At one TeV, the efficiency is flat out to 100~m.  

\section{Summary}
The discussion given in this paper represent a progress report, and it is
clear that we still have a significant way to go towards a final `product'.
On the other hand, it should be obvious that IACT systems provide a 
wealth of redundant information, both on individual air showers and
on average shower properties, and that they promise a qualitative step 
forward in the determination of $\gamma$-ray energy spectra. The power
of the HEGRA IACT system, combined with the large sample of about 30000
$\gamma$-ray events from Mkn 501, can be used to understand the properties
of the detection system at a level of precision which seemed impossible before.

\section{Acknowledgements}
The results reported here represent the work of the HEGRA IACT group; 
in particular the contributions by G.~Hermann and A. Konopelko should
be emphasized. The HEGRA experiment is supported by
the German Ministry for Research 
and Technology BMBF, and by the Spanish Research Council
CYCIT. We thank the Instituto
de Astrofisica de Canarias for the use of the site and
for providing excellent working conditions. 

\begin{refs}
\item Aharonian, F. et al., Astron. Astrophys., in press (1997); astro-ph/9706019.
\item Daum, A. et al., Astropart. Phys., in press (1997); astro-ph/9704098.
\item Frass, A. et al., subm. for publication (1997); astro-ph/9707304.
\item Konopelko, A. et al., Astropart. Phys. 4, 199 (1996).
\item Ulrich, M. et al., subm. for publication (1997); astro-ph/9708003.
\end{refs}

\end{document}